\documentclass[%
 reprint,
 amsmath,amssymb,
 aps,
prd,
]{revtex4-2}

\usepackage{graphicx}
\usepackage{dcolumn,amsmath}
\usepackage{bm,tikz-cd}
\def\mb{\mathbf}

\def\ie{\begin{equation}\begin{aligned}}
\def\fe{\end{aligned}\end{equation}}

\begin{document}

\title{Systematic approach to $\ell$-loop planar integrands from the classical equation of motion}

\author{Yi-Xiao Tao}
\email{taoyx21@mails.tsinghua.edu.cn}
\address{Department of Mathematical Sciences, Tsinghua University, Beijing 100084, China\\
Nordita, KTH Royal Institute of Technology and Stockholm University, Hannes Alf\'{v}ens v\"{a}g 12, SE-106 91 Stockholm, Sweden}

\begin{abstract}
In this paper, we present a recursive method for $\ell$-loop planar integrands in colored quantum field theories. We start with the classical equation of motion and then pick out the comb component, which will help us to define the loop kernels. Then we construct the $\ell$-loop integrands based on some recursion rules for the $\ell$-loop kernels. Finally, we reach a recursion formula for the $\ell$-loop planar integrands. Our method can be easily generalized to general quantum field theories, even non-Lagrangian theories, to obtain the planar part of the whole $\ell$-loop integrands.
\end{abstract}

\maketitle

\section{Introduction}
Scattering amplitudes are the most important objects in quantum field theory, which are highly relevant to high-energy experiments. Amplitudes have been studied for a very long time, and the calculation methods have improved a lot compared with the traditional Feynman rules, both numerically and analytically.

Most of the development on the analytic side comes from the success of the on-shell methods, see \cite{Elvang:2013cua,Travaglini:2022uwo} for some reviews. Besides the on-shell method, off-shell methods have also been developed in various fields \cite{Schubert:2001he}. Off-shell objects have a richer structure and can help us know more about different theories. In addition, off-shell objects are very useful for generating loop integrands or computing amplitudes recursively. One of the most successful off-shell objects is the Berends-Giele (BG) current \cite{Berends:1987me}. The BG current, which has only one off-shell leg, has many applications in different cases \cite{Mizera:2018jbh,Gomez:2021shh,Armstrong:2022mfr,Chattopadhyay:2021udc,Chattopadhyay:2024kdq,Wu:2021exa,Mafra:2010jq,Mafra:2015vca,Mafra:2016ltu,Tao:2023wls,Tao:2023yxy,Tao:2024vcz,Frost:2020eoa,Cho:2021nim,Naculich:2023wyp,Lopez-Arcos:2019hvg}. Hence, to develop off-shell techniques is very important both in theory and practical calculations.

Multi-particle solutions of the equation of motion, which is a method of classical field theory, can be regarded as an off-shell generalization of BG currents. Such objects are very useful in recursion. There are some examples of obtaining 1-loop integrands from the multi-particle solution of the classical equation of motion \cite{Gomez:2022dzk}. Actually, we can also obtain the multi-particle solution of a quantum version of the equation of motion directly \cite{Lee:2022aiu,Garg:2024icm}. Multi-particle solutions are more suitable for dealing with the loop integrands. We do not need to worry about the missing information because of the on-shell condition, since all external legs of the multi-particle solutions are off-shell.

In this paper, we want to develop techniques based on the multi-particle solutions. We start with the equation of motion of the bi-adjoint theory, and finally reach the whole $\ell$-loop planar integrand results, which correspond to the partial amplitudes associated with a single-trace structure in the color-trace decomposition \cite{Elvang:2013cua}. This systematic approach to $\ell$-loop planar integrands is based on the \text{comb component} of the multi-particle solutions and the $\ell$-\textit{loop kernel}. These two objects will be defined later. Based on these two objects, together with generalized BG currents, we will obtain the $\ell$-loop planar integrands recursively. We will also consider the Yang-Mills case and show some examples. It is worth mentioning that the $\ell$-loop planar integrands of the Yang-Mills theory can also be derived using planar variables \cite{Arkani-Hamed:2024tzl,Cao:2025mlt}. This method can be generalized to any theory to obtain the $\ell$-loop planar integrands by considering the planar component of the multi-particle solution to rule out the non-planar part. In this paper, the word `` $\ell$-loop integrands" is equivalent to ``$\ell$-loop planar integrands" since we will not discuss the non-planar ones.

This paper is organized as follows. We will focus on the bi-adjoint scalar theory and the Yang-Mills theory. We first consider the bi-adjoint scalar theory and define the comb component of the multi-particle solutions and the $\ell$-loop kernels. Then we will show how to reach the $\ell$-loop integrands of bi-adjoint theory using our systematic approach and show some examples. Finally, we will show how to apply our method to the Yang-Mills theory. We will choose some Feynman diagrams as examples and show that our method can reproduce them.

\section{Bi-adjoint: comb component and 1-loop kernel}\label{sec2}
Consider the bi-adjoint scalar theory with the following equation of motion:
\ie
\square \phi=\frac{1}{2}[[\phi,\phi]].
\fe
where $\phi=\phi_{a\tilde{a}} T^a\otimes\tilde{T}^{\tilde{a}}$. The notation $[[]]$ denotes the double commutator of two different generator $T^a$ and $\tilde{T}^{\tilde{a}}$, i.e.
\ie
[[T^a\otimes\tilde{T}^{\tilde{a}},T^b\otimes\tilde{T}^{\tilde{b}}]]=[T^a,T^b]\otimes[\tilde{T}^{\tilde{a}},\tilde{T}^{\tilde{b}}]
\fe
Using the perturbiner method \cite{Selivanov:1997aq,Selivanov:1998hn,Rosly:1997ap,Rosly:1998vm,Mizera:2018jbh}, we can obtain the multi-particle solution of the equation of motion above:
\ie
\phi&=\sum_{P,Q}\phi_{P|Q}e^{ik_P\cdot x}T^{a_P}\otimes\tilde{T}^{\tilde{a}_Q}\\
s_P\phi_{P|Q}&=\sum_{P=XY}\sum_{Q=WZ}(\phi_{X|W}\phi_{Y|Z}-\phi_{X|Z}\phi_{Y|W})
\fe
with $P=p_1\cdots p_m$, $T^{a_P}=T^{a_1}T^{a_2}\cdots T^{a_m}$, and $s_P=k_P^2=(\sum_{i=1}^mk_{p_i})^2$. Note that until now all external legs in $P$ are off-shell, which means that $k_i^2\neq0$ for $i\in P$. Now we define the following \textit{comb component} of the above solution, for simplicity we set $P=12\cdots m$:
\ie
s_P\phi^{\rm comb}_{12\cdots m|Q}=\sum_{Q=WZ}(\phi^{\rm comb}_{12\cdots m-1|W}\phi_{m|Z}-(W\leftrightarrow Z))
\fe
The above equation comes from the term $P=(12\cdots m-1,m)$ of the deconcatenation sum, and then keeps this choice in the recursion. Imposing $Q=P$, we obtain the following result:
\ie\label{ccc}
\phi^{\rm comb}_{12\cdots m|12\cdots m}=\frac{\prod_{i=1}^m\phi_{i|i}}{s_{1\cdots m}s_{1\cdots m-1}\cdots s_{12}}
\fe
Since all external legs are off-shell, we will keep the notation $\phi_{i|i}$ rather than using $\phi_{i|i}=1$ to remind this. For the bi-adjoint scalar case, we will always use \eqref{ccc} as the comb component.

The next step is to define an \textit{$m$-way 1-loop kernel}. We consider the comb component with the last propagator $s_{12\cdots m}\phi^{\rm comb}_{l12\cdots m|l12\cdots m}$, then using the sewing procedure \cite{Gomez:2022dzk} $\phi_{l|l}\to 1/l_1^2$ and imposing the momentum conservation for loop integrands, we can obtain the $m$-way 1-loop kernel
\ie
I^{\rm kernel}_{1,m}=\frac{\prod_{i=1}^m\phi_{i|i}(1-\frac{1}{2}\delta_{m,2})}{l_1^2(l_1+k_1)^2\cdots(l_1+\sum_{i=1}^{m-1}k_i)^2}
\fe
with the loop momentum $l_1$. Note that when $m=2$ we have a symmetry factor of 1/2. Consider a $n$-point 1-loop integrand $I^{\rm 1-loop}(12\cdots n|Q)$, we just need to divide $(12\cdots n)$ into $m$ parts, say $P_m$. Note that when constructing loop integrands, the ordered sets here are \textit{cyclic sets}, which means they are equivalent under cyclic permutation. As an example, the division we considered here also includes cases like $(n-1,n,1,2|3,4,\cdots,n-2)$. Then we do the following replacement
\ie\label{replace}
I^{\rm 1-loop}_m(12\cdots n|Q)=I^{\rm kernel}_{1,m}\bigg|_{k_i\to k_{P_i},\phi_{i|i}\to\Phi_{P_i|Q_i}}.
\fe
Here, $\{Q_i\}$ is a division of the cyclic set $Q$, i.e. $(Q_1|Q_2|\cdots|Q_m)$ is a division of $Q$ up to a cyclic permutation of $Q$. And $\Phi_{P_i|Q_i}$ here is the Berends-Giele current with all legs in $P_i$ on-shell comparing to the off-shell multi-particle solution $\phi_{P_i|Q_i}$. Note that we must find a division so that for every $i$, the elements in $Q_i$ must be the same in $P_i$, otherwise this term of the integrand will vanish. The whole $n$-point 1-loop integrands can be expressed as
\ie\label{1-loop}
I^{\rm 1-loop}(12\cdots n|Q)=\sum_{m=2}^{n}\sum_{\text{ $m$-division of $P$}}I^{\rm kernel}_{1,m}\bigg|_{k_i\to k_{P_i}\atop \phi_{i|i}\to\Phi_{P_i|Q_i}}
\fe
Note that when $m=1$, we will get a tadpole, which will be zero after integrating the loop momenta. Therefore, we start the sum with $m=2$. 

In the following, we will show an example of the 3-point 1-loop integrand $I(123|321)$. For 2-divisions, we have
\ie
(12|3),(1|23),(31|2),
\fe
for $(123)$ and the corresponding division for $(321)$ is
\ie
(21|3),(1|32),(13|2).
\fe
However, for 3-divisions, we only have $(1|2|3)$ for $(123)$, but for $(321)$ we only have $(1|3|2)$, which cannot match with $(1|2|3)$ up to a cyclic permutation. Thus this case will vanish. Hence the final result is
\ie
I^{\rm 1-loop}(123|321)=&-\frac{1}{2l^2(l+k_1)^2s_{23}}-\frac{1}{2l^2(l+k_{12})^2s_{12}}\\
&-\frac{1}{2l^2(l+k_{31})^2s_{31}}
\fe

\section{Graph factor}\label{sec3}
Before we go to the $\ell$-loop case, we need to define the \textit{graph factor} here in order to avoid overcounting. For an $\ell$-loop Feynman diagram, a graph factor can be written as $g=S\times \frac{1}{\ell-r}$. Here, $S$ is the symmetry factor coming from the symmetry of the Feynman diagram, and $r$ is defined as below. We first define the largest loop to be the loop that is connected with all external legs and no other loop outside it. Then we consider all loops that are independent in this largest loop, i.e., the regions of different loops will not overlap. After choosing the largest loop, we will ignore the external legs and the 3-point vertices connected to them and regard the diagram just as a vacuum diagram. Then, each loop will have some propagators in common with the largest loop. If there is 1 common propagator, we assign the loop a value of 0; otherwise 1. This value is called the \textit{loop value}. Then $r$ is the sum of the values of all loops.

In addition, there is an extra contribution to the symmetry factor. For 2-way kernels, there is a symmetry under flipping the diagram. By the word ``flip", we mean ``turn upside down". This will lead to some overcounts when we do the recursion introduced in the next section. Hence, we will regard the 2-way kernel and its flipping as two different diagrams when recursion, and these two diagrams will both get an extra symmetry factor of 1/2, unless the 2-way kernel and its flipping coincide. We will see an example later. Actually, in the bi-adjoint scalar case, all symmetry factors come from the 2-way kernel parts of the Feynman diagrams. Every 2-way kernel part contributes a factor of 1/2. A more elegant way to deal with it in the bi-adjoint theory has been demonstrated in \cite{Tao:2025pnt}.

The loop kernel without the symmetry factor $S$ in each diagram is called a \textit{bare loop kernel}. This concept is important when considering recursion, since when we go to higher-loop cases, the symmetry factor will change; hence, we only need to consider the symmetry factor in the last recursion. However, the factor $\frac{1}{\ell-r}$ here is to avoid overcounting in the following recursion; hence, we must consider it in every recursion.

\section{Bi-adjoint: Towards $\ell$-loop integrands}\label{sec4}
The $\ell$-loop kernel can be obtained from the $(\ell-1)$-loop kernel recursively. For a $m_\ell$-way $\ell$-loop kernel, we have the following steps to construct it
\begin{enumerate}
\item Consider ordered cyclic set $(b_\ell ,a_\ell,1,2, \cdots ,m_\ell-1,m_\ell)$ and the cyclic permutation of $(12\cdots m_\ell)$, i.e. there are $m_\ell$ sets in total. Note that due to the cyclic properties, $(1,2,\cdots,m_\ell-1,m_\ell,a_\ell,b_\ell)$ is equivalent to $(a_\ell,b_\ell,1,2\cdots ,m_\ell-1,m_\ell)$. 
\item Then, for $k$-divisions, which means we divide a set into $k$ parts, of each set, we set: i) $b_{\ell}$ itself to be one part of the $k$-division, ii) For a part not involving $a_{\ell}$, there can only be 1 element in the part, like $(b_{\ell}|a_{\ell},1|2|3)$. There is only 1 case for a given $k$ and a given set. 
\item Consider a $k$-way bare $(\ell-1)$-loop kernel, replace $\phi_{i|i}$ of this kernel with the comb component $\phi^{\rm comb}$ of each part of a $k$-division, just similar to \eqref{replace}. Then replace $\phi_{a_\ell|a_\ell}\phi_{b_\ell|b_\ell}$ with $1/l_\ell^2$ and $k_{a_\ell}=-k_{b_\ell}=l_\ell$. This replacement is equivalent to turning legs $a_\ell$ and $b_\ell$ into an internal line. Then add the corresponding graph factor $g$ to each case. Since we have demonstrated the construction of the Feynman diagrams, one can easily draw diagrams according to the steps above and find the corresponding graph factors.
\item Sum over all possible $k$-division and all $k\geq 2$, just like \eqref{1-loop}. Finally we will obtain a $m_\ell$-way $\ell$-loop kernel from $(\ell-1)$-loop kernel $I^{\rm kernel}_{\ell,m_\ell}$.
\end{enumerate}
From this recursion, one will find that $\ell$-loop kernels cannot be reduced to fewer loop diagrams by cutting a single propagator. It is important to explain the factor $r$ we defined before, after we give the rules for the recursion. In this recursion, a new loop is generated at the outside of the $(\ell-1)$-loop kernel, hence it must be part of the largest loop for the $\ell$-loop diagram. If the value of a loop is 0, then we can generate it through the recursion above. If the value of a loop is 1, there are two cases: it cannot be obtained by sewing, or it does not come from the $(\ell-1)$-loop kernel but a reducible $(\ell-1)$-loop diagram. In both cases, this loop cannot be generated by the recursion above.

\begin{figure}
	\centering
    \includegraphics[width=0.50\textwidth]{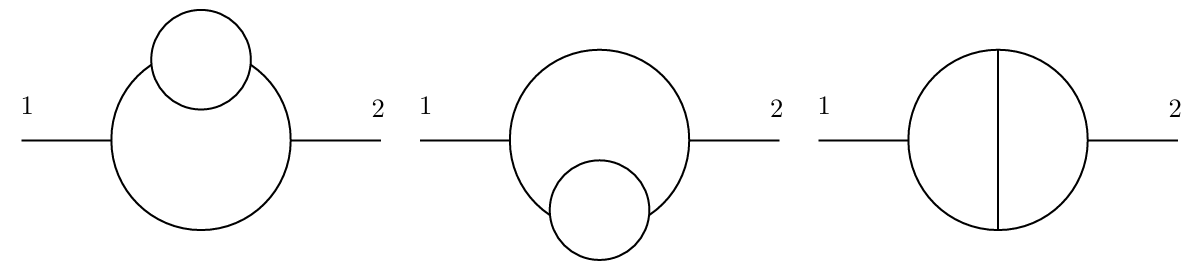}
	\caption{All diagrams in the 2-way 2-loop kernel}
	\label{scalar}
\end{figure}

Here we give an example of a 2-way 2-loop kernel. In this case, we need to consider $(b_2,a_2,1,2)$ and $(b_2,a_2,2,1)$ and their 2,3,4-divisions. For 2-division terms, we have 
\ie
\frac{\phi_{1|1}\phi_{2|2}}{l_1^2(l_1-l_2)^2l_2^4(l_2+k_1)^2}+\frac{\phi_{1|1}\phi_{2|2}}{l_1^2(l_1-l_2)^2l_2^4(l_2+k_2)^2}
\fe
For 3-division terms, we have
\ie
\frac{2\phi_{1|1}\phi_{2|2}}{l_1^2(l_1-l_2)^2(l_1+k_1)^2l_2^2(l_2+k_1)^2}
\fe
For 4-division terms, we have
\ie
\frac{\phi_{1|1}\phi_{2|2}}{l_1^4(l_1-l_2)^2(l_1+k_1)^2l_2^2}+\frac{\phi_{1|1}\phi_{2|2}}{l_1^4(l_1-l_2)^2(l_1+k_2)^2l_2^2}
\fe
We should add the symmetry factors to each term due to the automorphism of these diagrams. The final result is
\ie
I^{\rm kernel}_{2,2}=&\frac{\phi_{1|1}\phi_{2|2}}{2l_1^2(l_1-l_2)^2l_2^4(l_2+k_2)^2}\\
&+\frac{\phi_{1|1}\phi_{2|2}}{2l_1^2(l_1-l_2)^2(l_1+k_1)^2l_2^2(l_2+k_1)^2},
\fe
after some shifts of loop momenta. Note that we add an extra symmetry factor of 1/2 to the first term compared to the second term for the reason mentioned in Section \ref{sec3}. The first term corresponds to the first 2 diagrams of FIG. \ref{scalar}, which comes from the 2-division and the 4-division terms. The loops in these two diagrams are not invariant under flipping, while the loops in the third diagram, which corresponds to the second term and comes from the 3-division terms, are. It is also easy to find that all these diagrams have $r=0$.

We can construct $\ell$-loop integrands from the $\ell$-loop kernel, but this will only include the case where all loops are combined; i.e., the irreducible part. This part cannot be reduced to 2 diagrams with fewer loops by cutting a single propagator. In order to include the reducible part, let us first define the $\ell$-loop current $\Phi^{(\ell)}_{P|Q}$ with $\Phi^{(0)}_{P|Q}=\Phi_{P|Q}$:
\ie
\lim_{\rm on-shell}k_P^2\Phi^{(\ell)}_{P|Q}\phi_{l|l}=I^{\rm \ell-loop}(Pl|Ql)
\fe
which means that for bi-adjoint scalars we have
\ie
\Phi^{(\ell)}_{P|Q}=\frac{1}{k_P^2}I^{\rm \ell-loop}(Pl|Ql)\bigg|_{k_l\to -k_P}.
\fe
Then the final result, which includes both parts, is
\ie\label{n-loop}
&I^{\rm \ell-loop}(12\cdots n|Q)=\sum_{k=2}^{n}\sum_{\text{ $k$-division of $P$}}I^{\rm kernel}_{\ell,k}\bigg|_{k_i\to k_{P_i}\atop \phi_{i|i}\to\Phi_{P_i|Q_i}}\\
&+\frac{1}{\ell}(\sum_{m=1}^{\ell-1}\sum_{k=2}^{n}\sum_{\ell_i\atop\sum_{i=1}^k \ell_i=\ell-m}\sum_{\text{ $k$-division of $P$}}mI^{\rm kernel}_{m,k}\bigg|_{k_i\to k_{P_i}\atop \phi_{i|i}\to\Phi^{(\ell_i)}_{P_i|Q_i}})
\fe
with $P=123\cdots n$. Note that the factor $m/\ell$ in the second line is very important to avoid overcounting. Consider a given reducible diagram, it has $n_m$ number of $m$-loop parts and it satisfies $\sum_{m=1}^{\ell-1}m\times n_m=\ell$. For the $m$-loop kernel term in the second line of \eqref{n-loop}, there will be $n_m$ number of overcounts since every $m$-loop part is equivalent. After adding the factor $m/\ell$, the total counting is $\sum_{m=1}^{\ell-1}n_m\times m/\ell=1$. Hence we have canceled all the overcounts. The symmetry factors will not bother us in \eqref{n-loop}. These factors only affect the recursion of the loop kernel. Adding $\ell$-loop current $\Phi^{(\ell)}_{P|Q}$ to loop kernels does nothing to the symmetry factor.

Here we will show a 2-point 2-loop example using our formula \eqref{n-loop}. We have
\ie
&I^{\rm 2-loop}(12|12)=\sum_{\text{ $2$-division of $(12)$}}I^{\rm kernel}_{2,2}\bigg|_{k_i\to k_{P_i}\atop \phi_{i|i}\to\Phi_{P_i|Q_i}}\\
&+\frac{1}{2}(\sum_{\ell_1,\ell_2\atop\ell_1+\ell_2=1}\sum_{\text{ $2$-division of $(12)$}}I^{\rm kernel}_{1,2}\bigg|_{k_i\to k_{P_i}\atop \phi_{i|i}\to\Phi^{(\ell_i)}_{P_i|Q_i}})
\fe
There is only one case, $(1|2)$, in the $2$-division of $(12)$. Then we have 
\ie
&I^{\rm 2-loop}(12|12)=I^{\rm kernel}_{2,2}\bigg|_{\phi_{i|i}\to1}\\
&+\frac{1}{2}(I^{\rm kernel}_{1,2}\bigg|_{\phi_{1|1}\to\Phi^{(1)}_{1|1}\atop\phi_{2|2}\to 1}+I^{\rm kernel}_{1,2}\bigg|_{\phi_{2|2}\to\Phi^{(1)}_{2|2}\atop\phi_{1|1}\to 1})\\
&=\frac{1}{2l_1^2(l_1-l_2)^2(l_1+k_1)^2l_2^2(l_2+k_1)^2}\\
&+\frac{1}{2l_1^2(l_1-l_2)^2l_2^4(l_2+k_2)^2}+\frac{1}{4l_1(l_1+k_1)^2k_2^2l_2(l_2+k_1)^2}
\fe
Note that a 2-way 1-loop kernel has a symmetry factor of 1/2, which is the origin of $1/4$ in the last term. We also give some more examples in the appendix.

\section{Yang-Mills: A more general case}\label{sec5}
In the Yang-Mills case, the rules for loop kernel recursion and the process towards $\ell$-loop integrands are similar to those in the bi-adjoint case. We can also find the correct graph factors for the Yang-Mills Feynman diagrams just as before. Hence, we will not go into too much detail here. Instead, we will show some examples and calculations. In this section, we will show some examples of comb components and loop kernels. Then we will choose some 2-loop Feynman diagrams and see how to reproduce them using our method. Note that we will not consider the ghost fields here. Some 1-loop results can be found in \cite{Chen:2023bji,Gomez:2022dzk,Cao:2024olg,Caron-Huot:2010fvq}.

We will use the Feynman gauge for the off-shell multi-particle solution of gluons. The classical equation of motion of the Yang-Mills theory is
\ie
\square \mb{A}_{\mu}=-i[\mb{A}^{\nu},\mb{F}_{\mu\nu}]+i\partial_{\nu}[\mb{A}^{\nu},\mb{A}_{\mu}]
\fe
with $\mb{F}_{\mu\nu}=\partial_{\mu}\mb{A}_{\nu}-\partial_{\nu}\mb{A}_{\mu}-i[\mb{A}_{\mu},\mb{A}_{\nu}]$. Using the perturbiner method, we have
\ie
\mb{A}_{\mu}&=\sum_{P}A_{P\mu}e^{ik_P\cdot x}T^{a_P}\\
k_{P}^{2}A_{P\mu}&=\sum_{P=XY}(k_{X\mu}-k_{Y\mu})(A_{X}\cdot A_{Y})-A_{X\mu}(k_{X}\cdot A_{Y})\\
&-A_{X\mu}(k_{P}\cdot A_{Y})+A_{Y\mu}(k_{Y}\cdot A_{X})+A_{Y\mu}(k_{P}\cdot A_{X})\\
&+\sum_{P=XYZ}2A_{Y\mu}(A_{X}\cdot A_{Z})-A_{X\mu}(A_{Y}\cdot A_{Z})\\
&-A_{Z\mu}(A_{X}\cdot A_{Y})
\fe
The comb component is defined as before. For the 3-deconcatenation sum $\sum_{P=XYZ}$, we regard this as $\sum_{P=XW}\sum_{W=YZ}$ and then implement the definition for comb component before. Here we give an example
\ie
&k_{l12}^2A_{l12\mu}^{\text{comb}}=(k_{l1\mu}-k_{2\mu})(A_{l1}\cdot A_{2})-A_{l1\mu}(k_{l1}\cdot A_{2})\\
&-A_{l1\mu}(k_{l12}\cdot A_{2})+A_{2\mu}(k_{2}\cdot A_{l1})+A_{2\mu}(k_{l12}\cdot A_{l1})\\
&+2A_{1\mu}(A_{l}\cdot A_{2})-A_{l\mu}(A_{1}\cdot A_{2})-A_{2\mu}(A_{l}\cdot A_{1})
\fe
where
\ie
&k_{l1}^{2}A_{l1\mu}=(k_{l\mu}-k_{1\mu})(A_{l}\cdot A_{1})-A_{l\mu}(k_{l}\cdot A_{1})\\
&-A_{l\mu}(k_{l1}\cdot A_{1})+A_{1\mu}(k_{1}\cdot A_{l})+A_{1\mu}(k_{l1}\cdot A_{l})
\fe
Based on this, we obtain the 2-way 1-loop kernel after the replacement $A_{l\mu}A_{3\nu}\to \eta_{\mu\nu}/l_1^2$ in $k_{l12}^2A_{l12}\cdot A_3$ \cite{Gomez:2022dzk}:
\begin{widetext}
\ie
&l^2_1I_{1,2}^{\text{kernel}}=\frac{1}{2(l_1+k_1)^2}\bigg((k_{l1}-k_{2})\cdot A_1(k_{l}-k_{1})\cdot A_2-(k_{l1}-k_{2})\cdot A_2(k_{l}\cdot A_{1})-(k_{l1}-k_{2})\cdot A_2(k_{l1}\cdot A_{1})\\
&+(k_{l1}-k_{2})\cdot k_1 A_1\cdot A_2+(k_{l1}-k_{2})\cdot k_{l1} A_1\cdot A_2-((k_{l}-k_{1})\cdot A_{1}-d(k_{l}\cdot A_{1})-d(k_{l1}\cdot A_{1})+A_{1}\cdot k_{1}+A_{1}\cdot k_{l1})(k_{l1}\cdot A_{2})\\
&-((k_{l}-k_{1})\cdot A_{1}-d(k_{l}\cdot A_{1})-d(k_{l1}\cdot A_{1})+A_{1}\cdot k_{1}+A_{1}\cdot k_{l1})(k_{l12}\cdot A_{2})+k_{2}\cdot (k_{l}-k_{1})(A_{2}\cdot A_{1})-k_2\cdot A_{2}(k_{l}\cdot A_{1})\\
&-k_2\cdot A_{2}(k_{l1}\cdot A_{1})+k_2\cdot A_{1}(k_{1}\cdot A_{2})+k_2\cdot A_{1}(k_{l1}\cdot A_{2})+k_{l12}\cdot (k_{l}-k_{1})(A_{2}\cdot A_{1})-k_{l12}\cdot A_{2}(k_{l}\cdot A_{1})-k_{l12}\cdot A_{2}(k_{l1}\cdot A_{1})\\
&+k_{l12}\cdot A_{1}(k_{1}\cdot A_{2})+k_{l12}\cdot A_{1}(k_{l1}\cdot A_{2})\bigg)+\frac{1}{2}A_{1}\cdot A_{2}-\frac{1}{2}d(A_{1}\cdot A_{2})
\fe
\end{widetext}
where $k_l=l_1$ and the factor of 1/2 is the symmetry factor.

Note that here the last two terms correspond to the loop with a 4-point vertex. For the higher-loop kernels, one can construct them using the method in section \ref{sec3}. There is another subtlety due to the appearance of 4-point vertices. For higher-way (more than 2) 1-loop kernels, if we do the same thing as the bi-adjoint theory case, i.e., consider $A^{\rm comb}_{l12\cdots m\mu}$ only, we will find that there must be a propagator between leg 1 and leg $m$. However, there also exists a case that leg 1 and leg $m$ share a common 4-vertex. The lack of this case will break the cyclic invariance of the 1-loop kernels. Hence, we need to add an extra term besides the comb component in order to keep the cyclic invariance of 1-loop kernels. This term, called the \textit{contact component}, has the same definition as the comb component but with the last deconcatenation being a 3-deconcatenation. For example, the contact component of $A_{1234\mu}$ is
\ie
&k_{1234}^2A_{1234\mu}^{\rm contact}=(k_{123\mu}-k_{4\mu})(A^{\rm contact}_{123}\cdot A_{4})\\
&-A^{\rm contact}_{123\mu}(k_{123}\cdot A_{4})-A^{\rm contact}_{123\mu}(k_{1234}\cdot A_{4})\\
&+A_{4\mu}(k_{4}\cdot A^{\rm contact}_{123})+A_{4\mu}(k_{1234}\cdot A^{\rm contact}_{123})\\
&k_{123}^2A^{\rm contact}_{123\mu}=2A_{2\mu}(A_{1}\cdot A_{3})-A_{1\mu}(A_{2}\cdot A_{3})\\
&-A_{3\mu}(A_{1}\cdot A_{2})
\fe
In the case $P=1234$, we have ignored all the 3-deconcatenation terms like $P=(12,3,4)$ since the last deconcatenation is a 2-deconcatenation like $12=(1,2)$. To construct $m$-way 1-loop kernels with $m>2$, we need to consider
\ie
A^{\rm comb}_{l12\cdots m\mu}+A^{\rm contact}_{lm12\cdots (m-1)\mu}
\fe
instead of only the comb component. Such contact components appear only when we construct the 1-loop kernels. When we go to higher-loop integrands recursively, we do not need to consider contact components anymore. After figuring out the graph factors similar to what we have done for the bi-adjoint scalars, a systematic approach to $\ell$ gluon-loop Yang-Mills integrands is established. Note that the contribution from ghost-loops can be restored by considering the equation of motion involving ghost fields \cite{Gomez:2022dzk}. 

\section{Yang-Mills: Examples}
\begin{figure}
	\centering
    \includegraphics[width=0.30\textwidth]{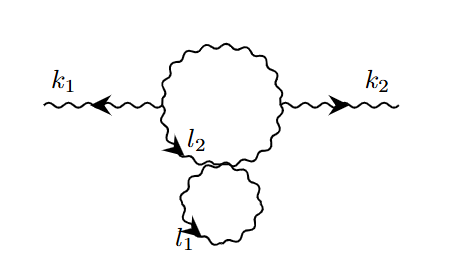}
	\caption{2-point 2-loop diagram 1}
	\label{fig1}
\end{figure}

In this section, we will give some examples of 2-loop YM planar integrands. Here, to simplify the calculation, we will just consider one of the possible divisions that can reach given Feynman diagrams and ignore some extra graph factors. We just aim to show that our method is consistent with the results from the Feynman rules. Consider the 2-point 2-loop diagram 1 (FIG. \ref{fig1}), we can generate this diagram by considering the last two terms of $I_{1,2}^{\text{kernel}}$ for the YM theory, and then consider the 2-division $(b_2|a_2,1,2)$ and the comb component $A^{\rm comb}_{a_212\mu}$. This will lead to the following result
\ie
&I^{\rm 2-loop}_{\rm FIG. 2}(12)=(1-d)\frac{(2l_2^2-2k_2\cdot l_2)+(4d-6)(l_{2}\cdot \epsilon_{1})(l_{2}\cdot \epsilon_{2})}{2l_1^2l_2^4(l_2+k_1)^2}.
\fe
Note that we have used the on-shell condition here. 

\begin{figure}
	\centering
    \includegraphics[width=0.30\textwidth]{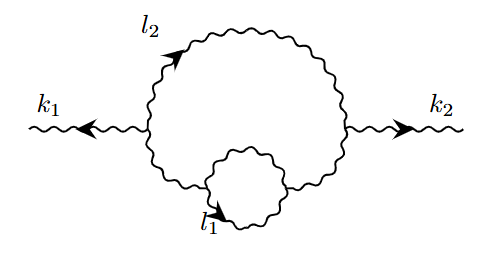}
	\caption{2-point 2-loop diagram 2 ($l_2$ here is the shifted one)}
	\label{fig2}
\end{figure}

For the 2-point 2-loop diagram 2 (FIG. \ref{fig2}), we consider the 2-division $(b_2|a_2,1,2)$ and the comb component $A^{\rm comb}_{a_212\mu}$ again, together with all but the last two terms of $I_{1,2}^{\rm kernel}$, to construct the corresponding off-shell diagram, and then take the on-shell limit. The final result is
\begin{widetext}
\ie
&2l_1^2(l_1+l_2+k_1)^2l_2^2(l_2+k_1)^2(l_2-k_2)^2I^{\rm 2-loop}_{\rm FIG. 3}(12)=\\
&\epsilon_1\cdot \epsilon_2 \bigg((4 d-6) \left(k_2\cdot l_1\right){}^2+2 k_2\cdot l_1 \left((4 d-6) l_1\cdot l_2+(2 d-5) l_2\cdot l_2+2 k_2\cdot l_2\right)-2 l_2\cdot l_2 \left((1-2 d) l_1\cdot l_2+15 k_2\cdot l_2-2 l_1\cdot l_1\right)\\
&+2 (2 d-3) \left(l_1\cdot l_2\right){}^2+(d+4) \left(l_2\cdot l_2\right){}^2+20 \left(k_2\cdot l_2\right){}^2-4 \left(l_1\cdot l_1+l_1\cdot l_2\right) k_2\cdot l_2\bigg)+2 (2 d-3) \epsilon_1\cdot l_1 \bigg(\epsilon_2\cdot l_1 \left(l_2\cdot l_2-4 k_2\cdot l_2\right)\\
&+\epsilon_2\cdot l_2 \left(3 k_2\cdot l_1+k_2\cdot l_2-3 l_1\cdot l_2-l_2\cdot l_2\right)\bigg)+\epsilon_1\cdot l_2 \bigg(\epsilon_2\cdot l_2 \left((19 d-24) l_2\cdot l_2-2 (2 d-3) \left(3 k_2\cdot l_1+10 k_2\cdot l_2-6 l_1\cdot l_1-3 l_1\cdot l_2\right)\right)\\
&+2 (2 d-3) \epsilon_2\cdot l_1 \left(3 k_2\cdot l_1+k_2\cdot l_2-3 l_1\cdot l_2-l_2\cdot l_2\right)\bigg)
\fe
\end{widetext}
Note that here we have done a shift $l_2\to -l_2-k_1$ so that the result is easy to compare with the result from FeynCalc. Both our results match the results from FeynCalc \cite{Hahn:2000kx}.


\section{Discussions}
In this paper, we give a systematic approach to $\ell$-loop planar integrands. We divide the $\ell$-loop planar integrands into 2 parts, the irreducible part and the reducible part. The irreducible part can be obtained from the $\ell$-loop kernel, which can be obtained recursively from the kernel with fewer loops, and some BG currents. The reducible part can be reached from the loop kernel with fewer loops and some generalized BG currents, which can be reached from loop integrands with fewer loops. Hence, the $\ell$-loop planar integrands can be determined by some objects with fewer loops, and all these constitute our recursion relation \eqref{n-loop}. 

This new recursion method, which starts with the classical equation of motion, is somehow a higher-loop version of the Berends-Giele recursion, which organizes the Feynman rules in a proper way to make the recursion properties manifest. In our $\ell$-loop recursion, we also creatively organized the Feynman rules so that we can reach $\ell$-loop integrands recursively. In contrast to other methods towards $\ell$-loop planar integrands, like \cite{Cao:2025mlt,Arkani-Hamed:2024tzl}, which can only be used in limited theories, our method has broader applications since it is based on the Feynman rules rather than some special properties of some specific theories. 

This powerful method has many potential applications. The most basic application is to construct the integrands in various theories, especially color theories. This method can also be used in non-Lagrangian theories since we only need the equation of motion \cite{Lyakhovich:2005mk}. Moreover, as a systematic approach, this method will help us to prove the relations between amplitudes of different theories, like \cite{Cheung:2017ems,Tao:2022nqc,Chen:2023bji}, since we can construct integrands in different theories with the same method.

An interesting thing to do in the future is to generalize this method to non-planar integrands. For example, we can generalize our method to the gravity case and look for some applications in the double copy relation at the loop level \cite{Bern:2008qj,Bern:2010ue,Correa:2024mub,Mazloumi:2024wys}.

\begin{acknowledgements}
YT would like to thank Qu Cao, Hadleigh Frost, and Zeyu Li for the discussions. YT also wants to thank Qu Cao for the valuable suggestions on the draft. Especially, YT thanks Chen Huang for helping draw the figures. YT is supported by the National Key R\&D Program of China (NO. 2020YFA0713000 and NO. 124B2094).

\end{acknowledgements}

\bibliographystyle{apsrev4-1}
\bibliography{planarloop}
\clearpage
\begin{widetext}
\appendix

\section{The 2-loop kernel with more ways}\label{app1}
In this appendix, we will give the 3-way and 4-way 2-loop kernels in detail. Let us consider the 3-way 2-loop kernel first. We will write down the division we need to consider and the corresponding terms:
\begin{enumerate}
\item The 2-division case
\ie
(b_2|a_2,1,2,3)+\text{cyclic}(1,2,3)
\fe
\ie
I^{(2)}_{2,3}=\frac{\phi_{1|1}\phi_{2|2}\phi_{3|3}}{l_1^2(l_1-l_2)^2l_2^4(l_2+k_1)^2(l_2+k_1+k_2)^2}+\text{cyclic}(k_1,k_2,k_3)
\fe
\item The 3-division case
\ie
(b_2|a_2,1,2|3)+\text{cyclic}(1,2,3)
\fe
\ie
I^{(3)}_{2,3}=\frac{\phi_{1|1}\phi_{2|2}\phi_{3|3}}{l_1^2(l_1-l_2)^2(l_1+k_1+k_2)^2l_2^2(l_2+k_1+k_2)^2(l_2+k_1)^2}+\text{cyclic}(k_1,k_2,k_3)
\fe
\item The 4-division case
\ie
(b_2|a_2,1|2|3)+\text{cyclic}(1,2,3)
\fe
\ie
I^{(4)}_{2,3}=\frac{\phi_{1|1}\phi_{2|2}\phi_{3|3}}{l_1^2(l_1-l_2)^2(l_1+k_1)^2(l_1+k_1+k_2)^2l_2^2(l_2+k_1)^2}+\text{cyclic}(k_1,k_2,k_3)
\fe
\item The 5-division case
\ie
(b_2|a_2|1|2|3)+\text{cyclic}(1,2,3)
\fe
\ie
I^{(5)}_{2,3}=\frac{\phi_{1|1}\phi_{2|2}\phi_{3|3}}{l_1^4(l_1-l_2)^2(l_1+k_1)^2(l_1+k_1+k_2)^2l_2^2}+\text{cyclic}(k_1,k_2,k_3)
\fe
\end{enumerate}
Then we have
\ie
&I^{\rm kernel}_{2,3}=\frac{1}{4}I^{(2)}_{2,3}+\frac{1}{2}I^{(3)}_{2,3}+\frac{1}{2}I^{(4)}_{2,3}+\frac{1}{4}I^{(5)}_{2,3},
\fe
where the extra factor of each term is the graph factor.  

Then we turn to the 4-way 2-loop kernel. 
\begin{enumerate}
\item The 2-division case
\ie
(b_2|a_2,1,2,3,4)+\text{cyclic}(1234)
\fe
\ie
I^{(2)}_{2,4}=\frac{\phi_{1|1}\phi_{2|2}\phi_{3|3}\phi_{4|4}}{l_1^2(l_1-l_2)^2l_2^4(l_2+k_1)^2(l_2+k_1+k_2)^2(l_2+k_1+k_2+k_3)^2}+\text{cyclic}(k_1,k_2,k_3,k_4)
\fe
\item The 3-division case
\ie
(b_2|a_2,1,2,3|4)+\text{cyclic}(1,2,3,4)
\fe
\ie
I^{(3)}_{2,4}=\frac{\phi_{1|1}\phi_{2|2}\phi_{3|3}\phi_{4|4}}{l_1^2(l_1-l_2)^2(l_1+k_1+k_2+k_3)^2l_2^2(l_2+k_1)^2(l_2+k_1+k_2)^2(l_2+k_1+k_2+k_3)^2}+\text{cyclic}(k_1,k_2,k_3,k_4)
\fe
\item The 4-division case
\ie
(b_2|a_2,1,2|3|4)+\text{cyclic}(1,2,3,4)
\fe
\ie
I^{(4)}_{2,4}=\frac{\phi_{1|1}\phi_{2|2}\phi_{3|3}\phi_{4|4}}{l_1^2(l_1-l_2)^2(l_1+k_1+k_2)^2(l_1+k_1+k_2+k_3)^2l_2^2(l_2+k_1)^2(l_2+k_1+k_2)^2}+\text{cyclic}(k_1,k_2,k_3,k_4)
\fe
\item The 5-division case
\ie
(b_2|a_2,1|2|3|4)+\text{cyclic}(1,2,3,4)
\fe
\ie
I^{(5)}_{2,4}=\frac{\phi_{1|1}\phi_{2|2}\phi_{3|3}\phi_{4|4}}{l_1^2(l_1-l_2)^2(l_1+k_1)^2(l_1+k_1+k_2)^2(l_1+k_1+k_2+k_3)^2l_2^2(l_2+k_1)^2}+\text{cyclic}(k_1,k_2,k_3,k_4)
\fe
\item The 6-division case
\ie
(b_2|a_2|1|2|3|4)+\text{cyclic}(1,2,3,4)
\fe
\ie
I^{(6)}_{2,4}=\frac{\phi_{1|1}\phi_{2|2}\phi_{3|3}\phi_{4|4}}{l_1^4(l_1-l_2)^2(l_1+k_1)^2(l_1+k_1+k_2)^2(l_1+k_1+k_2+k_3)^2l_2^2}+\text{cyclic}(k_1,k_2,k_3,k_4)
\fe
\end{enumerate}
Then we have
\ie
I^{\rm kernel}_{2,4}=\frac{1}{4}I^{(2)}_{2,4}+\frac{1}{2}I^{(3)}_{2,4}+\frac{1}{2}I^{(4)}_{2,4}+\frac{1}{2}I^{(5)}_{2,4}+\frac{1}{4}I^{(6)}_{2,4}
\fe

As an application, the 3-point 2-loop planar integrand can be expressed using the elements above
\ie
&I^{\rm 2-loop}(123|123)=I^{\rm kernel}_{2,2}\bigg|_{k_2\to k_{23}\atop \phi_{1,1}\to1,\ \phi_{2|2}\to\Phi_{23|23}}+I^{\rm kernel}_{2,2}\bigg|_{k_1\to k_{31}\atop \phi_{2,2}\to1,\ \phi_{1|1}\to\Phi_{31|31}}+I^{\rm kernel}_{2,2}\bigg|_{k_1\to k_{12},\ k_2\to k_3\atop \phi_{2,2}\to1,\ \phi_{1|1}\to\Phi_{12|12}}+I^{\rm kernel}_{2,3}\bigg|_{\phi_{i|i}\to1}\\
&+\frac{1}{2}\bigg(I^{\rm kernel}_{1,2}\bigg|_{k_1\to k_{31}\atop \phi_{2|2}\to 1,\ \phi_{1|1}\to\Phi^{(1)}_{31|31}}+I^{\rm kernel}_{1,2}\bigg|_{k_2\to k_{23}\atop \phi_{1|1}\to 1,\ \phi_{2|2}\to\Phi^{(1)}_{23|23}}+I^{\rm kernel}_{1,2}\bigg|_{k_1\to k_{12},\ k_2\to k_3\atop \phi_{3|3}\to 1,\ \phi_{1|1}\to\Phi^{(1)}_{12|12}}+I^{\rm kernel}_{1,2}\bigg|_{k_1\to k_{31}\atop \phi_{2|2}\to \Phi^{(1)}_{2|2},\ \phi_{1|1}\to\Phi_{31|31}}\\
&+I^{\rm kernel}_{1,2}\bigg|_{k_2\to k_{23}\atop \phi_{1|1}\to \Phi^{(1)}_{1|1},\ \phi_{2|2}\to\Phi_{23|23}}+I^{\rm kernel}_{1,2}\bigg|_{k_1\to k_{12},\ k_2\to k_3\atop \phi_{3|3}\to \Phi^{(1)}_{3|3},\ \phi_{1|1}\to\Phi_{12|12}}+I^{\rm kernel}_{1,3}\bigg|_{\phi_{2|2}\to1,\ \phi_{3|3}\to1\atop\phi_{1|1}\to \Phi^{(1)}_{1|1}}+I^{\rm kernel}_{1,3}\bigg|_{\phi_{1|1}\to1,\ \phi_{3|3}\to1\atop\phi_{2|2}\to \Phi^{(1)}_{2|2}}\\
&+I^{\rm kernel}_{1,3}\bigg|_{\phi_{1|1}\to1,\ \phi_{2|2}\to1\atop\phi_{3|3}\to \Phi^{(1)}_{3|3}}
\bigg)
\fe
where $\Phi^{(1)}_{1|1}$ and $\Phi^{(1)}_{12|12}$ can be obtained for the 2-point 1-loop integrand and the 3-point 1-loop integrand easily.

\end{widetext}

\end{document}